\documentclass[superscriptaddress,showpacs,aps,twocolumn,prb,floatfix]{revtex4}

\usepackage{amssymb}
\usepackage{amsmath}
\usepackage[dvips]{graphicx}
\usepackage{subfigure}
\usepackage{bm}
\usepackage{color}
\usepackage{gensymb}
\usepackage{wasysym}
\usepackage{xcolor,soul}
\usepackage{url}

\begin{document}
 
\title{Strain induced magnetic transition in CaMnO$_3$ ultrathin films}

\author{A. L\'opez Pedroso}
\affiliation{Laboratorio de Nanoestructuras Magn\'eticas y Dispositivos, Departamento de F\'{\i}sica de la Materia Condensada, Centro At\'omico Constituyentes, 1650 San Mart\'{\i}n, Pcia. de Buenos Aires, Argentina}
\affiliation{Instituto de Nanociencia y Nanotecnolog\'{\i}a (INN CNEA-CONICET), 1650 San Mart\'{\i}n, Argentina}
\affiliation{Consejo Nacional de Investigaciones Cient\'{\i}ficas y T\'ecnicas, C1425FQB Ciudad Aut\'onoma de Buenos Aires, Argentina.}

\author{M.A. Barral}

\affiliation{Instituto de Nanociencia y Nanotecnolog\'{\i}a (INN CNEA-CONICET), 1650 San Mart\'{\i}n, Argentina}
\affiliation{Consejo Nacional de Investigaciones Cient\'{\i}ficas y T\'ecnicas, C1425FQB Ciudad Aut\'onoma de Buenos Aires, Argentina.}
\affiliation{Departamento de F\'{\i}sica de la Materia Condensada, GIyA-CNEA, Avenida
General Paz 1499, (1650) San Mart\'{\i}n, Pcia. de Buenos Aires, Argentina}

\author{M.E. Graf}
\affiliation{Consejo Nacional de Investigaciones Cient\'ificas y T\'ecnicas, C1425FQB Ciudad Aut\'onoma de Buenos Aires, Argentina.}
\affiliation{Instituto de F\'{\i}sica Rosario (CONICET-UNR), Rosario, Argentina}
\affiliation{Facultad de Ciencias Exactas, Ingenier\'ia y Agrimensura, Universidad Nacional de Rosario (FCEIA-UNR)}

\author{A.M. Llois}

\affiliation{Instituto de Nanociencia y Nanotecnolog\'{\i}a (INN CNEA-CONICET), 1650 San Mart\'{\i}n, Argentina}
\affiliation{Consejo Nacional de Investigaciones Cient\'{\i}ficas y T\'ecnicas, C1425FQB Ciudad Aut\'onoma de Buenos Aires, Argentina.}
\affiliation{Departamento de F\'{\i}sica de la Materia Condensada, GIyA-CNEA, Avenida
General Paz 1499, (1650) San Mart\'{\i}n, Pcia. de Buenos Aires, Argentina}

\author{M.H. Aguirre}
\affiliation{Instituto de Nanociencia de Arag\'on and Instituto de Ciencia de Materiales de Arag\'on, Universidad de Zaragoza, E-50018 Zaragoza, Spain.}
\affiliation{Departamento de F\'{\i}sica de la Materia Condensada, Universidad de Zaragoza, E-50009 Zaragoza, Spain.}
\affiliation{Laboratorio de Microscop\'{\i}as Avanzadas, Universidad de Zaragoza, E-50018 Zaragoza, Spain.}

\author{L.B. Steren}
\affiliation{Laboratorio de Nanoestructuras Magn\'eticas y Dispositivos, Departamento de F\'{\i}sica de la Materia Condensada, Centro At\'omico Constituyentes, 1650 San Mart\'{\i}n, Pcia. de Buenos Aires, Argentina}
\affiliation{Instituto de Nanociencia y Nanotecnolog\'{\i}a (INN CNEA-CONICET), 1650 San Mart\'{\i}n, Argentina}
\affiliation{Consejo Nacional de Investigaciones Cient\'{\i}ficas y T\'ecnicas, C1425FQB Ciudad Aut\'onoma de Buenos Aires, Argentina.}

\author{S. Di Napoli}

\affiliation{Instituto de Nanociencia y Nanotecnolog\'{\i}a (INN CNEA-CONICET), 1650 San Mart\'{\i}n, Argentina}
\affiliation{Consejo Nacional de Investigaciones Cient\'{\i}ficas y T\'ecnicas, C1425FQB Ciudad Aut\'onoma de Buenos Aires, Argentina.}
\affiliation{Departamento de F\'{\i}sica de la Materia Condensada, GIyA-CNEA, Avenida
General Paz 1499, (1650) San Mart\'{\i}n, Pcia. de Buenos Aires, Argentina}
\email{dinapoli@tandar.cnea.gov.ar}

\begin{abstract}
The effect of high tensile strain and low dimensionality on the magnetic and electronic properties of CaMnO$_3$ ultrathin films, epitaxially grown on SrTiO$_3$ substrates, are experimentally studied and theoretically analyzed. By means of \textit{ab initio} calculations, we find that, both, the high strain produced by the substrate and the presence of the free surface contribute to the stabilization of an in-plane ferromagnetic coupling, giving rise to a non-zero net magnetic moment in the ultrathin films. Coupled with this change in the magnetic order we find an insulator-metal transition triggered by the quantum confinement and the tensile epitaxial strain. Accordingly, our magnetic measurements in 3\,nm ultrathin films show a ferromagnetic hysteresis loop, absent in the bulk compound due to its G-type antiferromagnetic structure.
\end{abstract}

\maketitle

\section{Introduction}\label{intro}
\indent

In the last few years much attention has been devoted to antiferromagnetic materials in view of future technological applications.~\cite{RevModPhys2018, NatPhys2018,NatNanotch2016} Antiferromagnetic materials produce no demagnetization fields, are robust against magnetic field perturbations and show ultrafast dynamics with large magnetotransport responses. All these properties make them suitable to be considered as future candidates for spintronic devices. However, the use of antiferromagnetic materials in microelectronic devices has been elusive, due to the difficulty in manipulating and detecting antiferromagnetic spin arrangements, until the recent experimental achievement of  electrical switching of an antiferromagnetic CuMnAs thin-film by P. Wadley \textit{et al.} in 2016.~\cite{Wadley2016}\\
In the search after suitable materials for antiferromagnetic spintronics, consideration of transition metal perovskites is a must, as they are characterized by their strong coupling among structural, orbital, charge and spin degrees of freedom.  Within this class of materials, particular attention has to be paid to manganites, where the exchange interactions can be regarded as a competition between ferromagnetic double-exchange (DE)~\cite{Zener1951,deGennes1960} and antiferromagnetic superexchange interactions (SE).~\cite{KRAMERS1934,Goodenough1955} The magnetic interaction among Mn ions is very sensitive to their local atomic environment and, due to this sensitivity, lattice distortions imposed by epitaxial strain or low dimensionality can introduce strong changes in the properties of thin-film manganites. \\
Among the manganites, the perovskite oxide CaMnO$_3$ (CMO) is a promising material for many technological applications, it is a high-temperature thermoelectric material~\cite{Thiel2015} and could show novel transport phenomena as it can display non trivial spin textures that are tunable through doping, it is sensitive to  strain and suitable for interface engineering. \\ 
In the last decades, the role of electron doping in CMO, through the substitution of the divalent Ca atoms by trivalent ions $R^{3+}$ ($R$=La, Pr, Sm, Gd or Ce), has been experimentally and theoretically studied and it is nowadays well known that electron doping produces changes in its magnetic structure.~\cite{La-doped1, La-doped2, La-doped3} Bulk CMO has a G-type antiferromagnetic (GAF) order below T$_N$ = 120 K~\cite{pr_macchesney,Zeng1999,Loshkareva2004} with an additional weak ferromagnetic component in its ground state.\,\cite{Spaldin2011,Bibes2019} The Mn ions are in the Mn$^{4+}$ valence state, with their three remaining $3d$ electrons occupying the t$_{2g}$-orbitals. When CMO is electron-doped, this weak FM component increases and two distinct scenarios have been proposed for the origin of this enhancement: (i) Phase separation, where the GAF and FM domains coexist, but which is not consistent with the appearance of a metallic behaviour in some cases of electron-doping and (ii) a higher spin-canting in the GAF which deviates from the antiparallel alignment and induces a bigger FM component in order to gain energy through the hopping of doping electrons via a DE interaction.  \\
CaMnO$_3$ has also been used as a paradigmatic example of a magnetic perovskite in many previous studies.~\cite{Bibes2019-2,apl_takahashi,apl_imbrenda} For instance, the role of oxygen vacancies in the magnetic properties of bulk CMO under strain has  been the focus of previous works.~\cite{nanolett_chandrasena,Spaldin2013,ovacancy2014}
Furthermore, a large topological Hall effect has been recently found in thin films of slighly electron-doped CaMnO$_3$ through Ce$^{4+}$.~\cite{Bibes2019} \\
Bulk CaMnO$_3$ crystallizes in the orthorhombic perovskite structure having $Pnma$ space group with lattice parameters $a=5.28$~\AA, $b=5.27$~\AA~ and $c=7.47$~\AA.~\cite{pr_macchesney,Zeng1999,Zhou2006} This structure is related to the cubic perovskite by an out of plane tilt of the BO$_6$ octahedra around two of the cubic coordinate axes and an in plane tilt around the third axis, ($a^-a^-c^+$) in Glazer notation. The orthorhombic crystal structure of CMO can also be described as $\sqrt{2}a_p \cdot \sqrt{2}a_p \cdot 2a_p$ in the pseudo-cubic approximation with a lattice parameter $a_p=3.73$~\AA. \\
Due to their high versatility, manganites in general and CMO in particular belong to those promising candidates for antiferromagnetic spintronic devices. However, because of the strong coupling between the different degrees of freedom present in these compounds, a thorough study of the correlation among the different stimuli and their incidence on the physical properties are to be studied.  \\
Even if  CaMnO$_3$ has been the object of study in many previous theoretical works, most of them are focused in bulk properties and only a few works are pointing towards the low dimensional properties.~\cite{Picket1999,Nordstrom2017} In the present work we computationally investigate the effects of a strong tensile strain and low dimensionality on the magnetic and electronic properties of ultrathin CaMnO$_3$ films. By means of the calculations we find that the G-type antiferromagnetic structure of bulk CMO changes to the A-type antiferromagnetic configuration when an ultrathin CMO film is under high tensile strain. This would lead to the appearance of a ferromagnetic signal in the magnetic measurements of CaMnO$_3$ thin films, coming from non-compensated ferromagnetic MnO$_2$ layers. This FM component is not a consequence of a chemical electron-doping of the films, as in previous works, but rather a self-doping of the surface allowed by the quantum confinement and the absence of apical oxygens. In order to corroborate this result it is necessary to epitaxially grow CMO thin films on a substrate with a larger lattice parameter in such a way that the film should grow naturally strained. An appropiate candidate suitable for this purpose is an SrTiO$_3$ substrate, which crystallizes as a cubic perosvkite with a lattice parameter of $a_{STO}=3.905$\,\AA. Therefore, we compare our \textit{ab initio} results with the outcome of magnetic and structural measurements performed on  3nm-thick CMO films grown on SrTiO$_3$.

\section{The perfect C\MakeLowercase{a}M\MakeLowercase{n}O$_3$ strained ultrathin film} 
\label{Comp-details}
\begin{figure}
\includegraphics[width=.6\columnwidth]{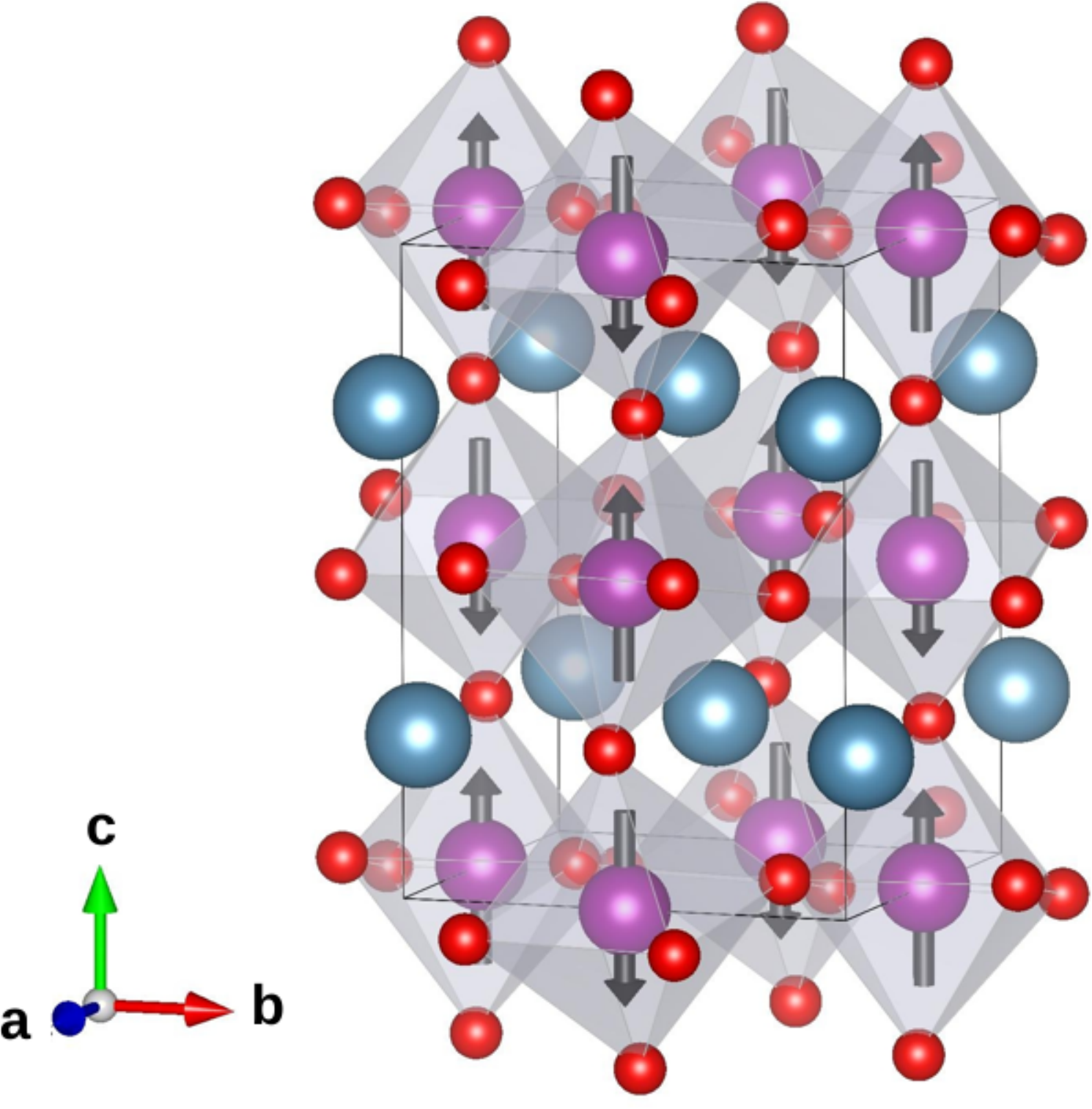}
\caption{(Color online) Schematic representation of CaMnO$_3$ \textit{Pnma} structure with the GAF magnetic configuration. Light blue: Ca, magenta: Mn and red: O atoms.}
\label{structure}
\end{figure}

The most important effects to take into account when dealing with the electronic and magnetic properties of thin films grown on top of substrates are: (i) strain effects, as thin films are usually epitaxially grown on top of a thick substrate, that imposes its in-plane lattice constant and symmetry and (ii) low dimensionality effects, as at the surface some of the interactions are missing due to the lack of neighboring atoms. The smaller the nanostructure, the more important the surface effects due to the increasing surface/volume ratio. In order to 
separate both effects we perform \textit{ab initio} calculations within  the framework of Density Functional Theory (DFT) and the projector  augmented wave (PAW) method,~\cite{PAW} as implemented in the Vienna \textit{ab initio} package (VASP).~\cite{VASP,PAW-VASP} We explicitly treat 10 valence electrons for Ca (3s$^2$3p$^6$4s$^2$), 13 for Mn (3p$^6$3d$^5$4s$^2$)  and 6 for  O (2s$^2$2p$^4$). The local spin density approximation (LSDA) in the parametrization of Ceperley and Alder is used.~\cite{LDA1,LDA2}  All the DFT calculations are performed using a 520~eV energy cutoff in the plane waves basis. As already shown, for manganese perovskites the local spin density approximation successfully predicts the observed stable magnetic phase and the structural parameters.~\cite{Picket1999, Picket2000, Nordstrom2017} We include a Hubbard term with $U=5$~eV and $J=1$~eV within the Lichtenstein implementation,~\cite{Liechtenstein95} for a better treatment of the Mn $3d$-electrons in CaMnO$_3$. With these parameters, the magnetic ground state is well reproduced.\\
The noncollinearity of the Mn magnetic moments is known to be quite minimal and gives rise to a small magnetic moment of 0.04$\mu_B$\,\cite{Spaldin2011,Bibes2019} and, therefore we approximate the magnetic configuration with a collinear model. The most relevant antiferromagnetic (AFM)  orders that might be shown by the Mn atoms are GAF (AFM coupling between first nearest neighbours), AAF (ferromagnetic (FM) coupling within the (001) planes and AFM between adjacent planes) and CAF (AFM within the (001) planes and FM between adjacent planes).
For bulk calculations we use a 20-atoms unit cell compatible with both the $Pnma$ crystallographic structure, as sketched in Fig.~\ref{structure}, and the four possible magnetic spin orientations. For the slab calculations we use a supercell that contains eight layers of CaO alternated with nine layers of MnO$_2$ along the (001) direction and a vaccum space of 15\AA~ to avoid the fictitious interaction between slabs due to the periodic boundary conditions. This number of layers leads to a slab width of around 2.84\,nm. To evaluate the integrals within the BZ a 6$\times$6$\times$6 and a 6$\times$6$\times$1 Monkhorst-Pack $k$-point grids are employed for bulk and slab calculations, respectively. The structural relaxations are performed until the forces on each ion are less than 0.01~eV/\AA.

\begin{figure}[ht!]
\includegraphics[width=.95\columnwidth]{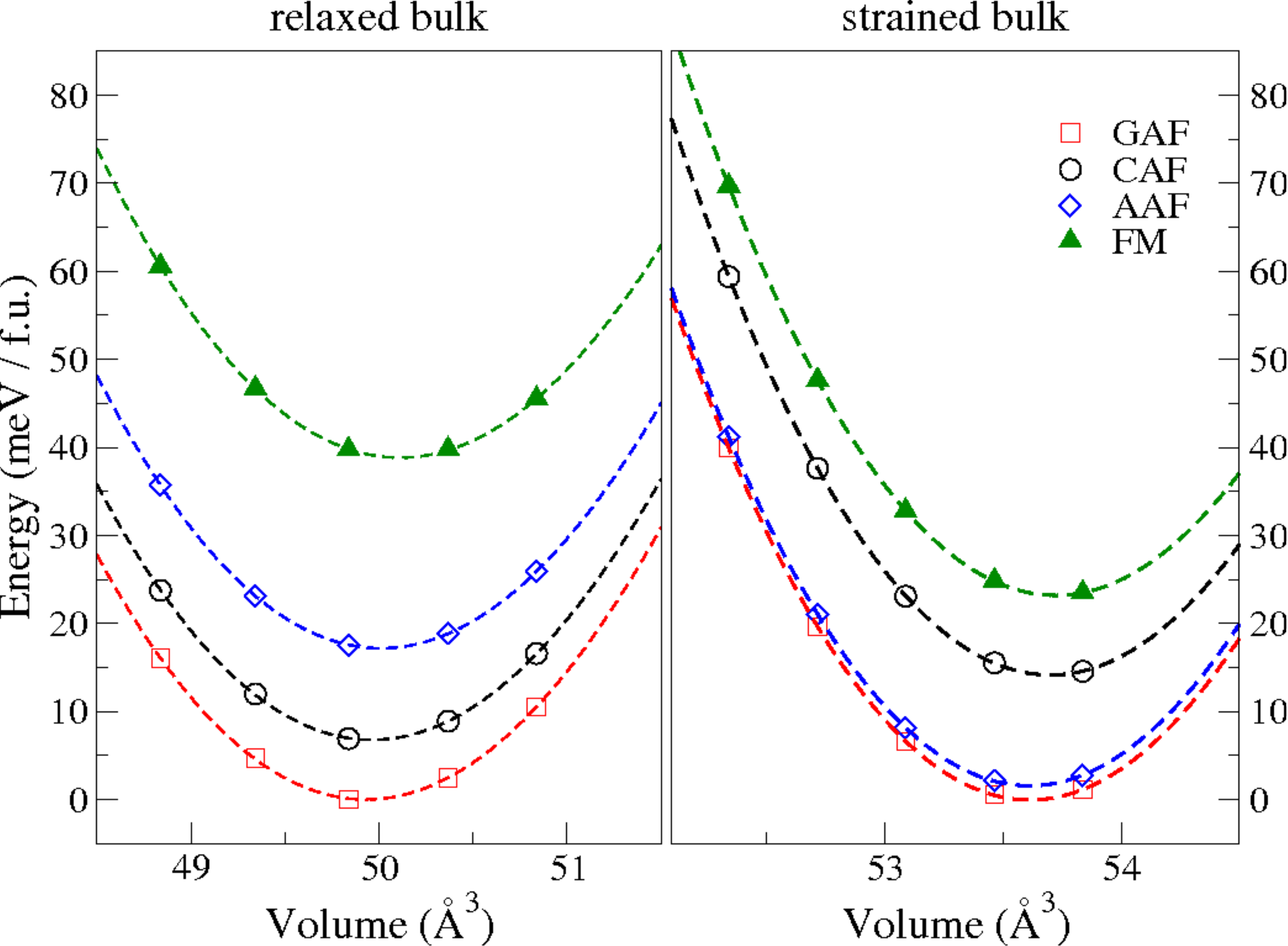}
\caption{(Color online) Energy difference versus volume curves for the different magnetic configurations in relaxed (left) and strained (right) bulk CaMnO$_3$. The continuous lines correspond to the fitted Birch-Murnaghan's equation of states for each case.}
\label{CMO-energy}
\end{figure}

As a first step, we study the stability of the mentioned  magnetic configurations as a function of volume for CaMnO$_3$ bulk. In the left panel of Fig.~\ref{CMO-energy} the total energies of the different magnetic structures relative to the ground state energy are plotted as a function of volume per formula unit. The data are fitted by the Birch-Murnaghan's equation of states.~\cite{Birch1947} It can be seen from this figure, that the most stable spin-ordered bulk phase corresponds to the GAF phase, in agreement with the literature, followed by the CAF, AAF and FM configurations, as already found in Refs.~\onlinecite{Spaldin2013} and \onlinecite{Terakura2010}. 

In Table~\ref{tab-CMO-bulk} we summarize the optimized lattice parameters, the band gap and the Mn's magnetic moment for all the studied cases. The obtained values agree well with the previous DFT calculations and, when it corresponds, with experimental values. 

\begin{table}[ht!]
\caption{Optimized lattice parameters in the $Pnma$ symmetry (\AA), band gap $\Delta_g$(eV) and Mn magnetic moment $m_{Mn}$ ($\mu_B$) for the different considered magnetic configurations in CaMnO$_3$ bulk.}
\label{tab-CMO-bulk}
\begin{tabular}{ccccccc}
\hline
			& $a$ 	& $b$ & $c$ & $\Delta_g$ & $m_{Mn}$ \\
\hline

Experimental~\cite{Zeng1999,Zhou2006,Paszkowicz,Loshkareva2004,Telegin2014}	                    & 5.280     & 5.270 	& 7.467 &    1.55 & 2.65   \\
GAF	            & 5.230     & 5.185     & 7.351 &	 1.60 & 2.71   \\
CAF             & 5.194     & 5.154     & 7.354 &    1.50 & 2.70   \\
AAF             & 5.241     & 5.171     & 7.282 &    1.40 & 2.72   \\
FM              & 5.233     & 5.168     & 7.332 &    1.25 & 2.76   \\
\hline\hline				
\end{tabular}
\end{table}

As the SrTiO$_3$ substrates induce a 4.7\% tensile epitaxial strain on CaMnO$_3$, we first analyze the effects of this strain on bulk CaMnO$_3$. In order to simulate this, we fix the lattice constants in the $ab$ plane to the corresponding of SrTiO$_3$ obtained by full optimization within the LDA, which is $a_{STO}^{LDA}=5.469$\AA~($a_{STO}=5.523$\AA) given in the orthorhombic symmetry. In the right panel of Fig.~\ref{CMO-energy} we plot the total energies of the different magnetic configurations as a function of volume per formula unit. 
The energy differences with respect to the ground state, magnetic moments and structural parameters for the four magnetic configurations in, both, the relaxed and the strained cases are given in Table~\ref{tab-CMO-structural}. We find that the energetic hierarchy in the strained CaMnO$_3$ changes with respect to the relaxed structure due to the induced tensile strain, which tends to make the AAF spin-order more stable than the CAF configuration (see also Fig.~\ref{CMO-energy}). As shown in Table~\ref{tab-CMO-structural} the bond-angles increase in the $ab$-plane and decrease along the $c$-axis, while the bond-lengths are larger in the $ab$-plane and shorter along the $c$-axis. As known, an increment in the bond-angle yields a stronger superexchange AFM coupling, while an increased bond-length leads to a smaller hopping strength, which tends to decrease the superexchange coupling. Our results show that the bond-length effect predominates and the in-plane ferromagnetic coupling tends to be more stable than the in-plane antiferromagnetic coupling, as observed in the energy hierarchy ($E_{AAF} < E_{CAF}$). As expected from previous works,~\cite{Terakura2010} the compound is insulating for all of the magnetic configurations, with the $t_{2g}\uparrow$ orbitals occupied, consistent with a Mn$^{4+}$ valence state and with a magnetic moment of around 3$\mu_B$ (see Table~\ref{tab-CMO-bulk}). The tensile strain breaks the degeneracies between the $e_g$ orbitals and the bottom of the conduction band has a $d_{x^2-y^2}$ character for all the magnetic structures. \\

\begin{table}[ht!]
\caption{For the relaxed and strained bulk: Energy with respect to the ground state magnetic configuration (in meV/f.u.), Mn magnetic moments, m$_{Mn}$ (in $\mu_B$), mean Mn-Mn, Mn-O bond-lengths (in \AA) and Mn-O-Mn bond-angles (in deg) in the in-plane and -out-of-plane directions for the different magnetic configurations. $\angle_{\parallel}$: in-plane Mn-O-Mn bond-angle. $\angle_{\perp}$: Mn-O-Mn bond-angle in the $c$-direction.}
\centering
\begin{tabular}{lcccc}
\hline
\hline
&&\textbf{Relaxed bulk} &&\\
\hline
Magnetic order	& GAF& CAF & AAF & FM  \\
\hline \hline
$\Delta E$  & 0 & 7 & 17 & 39\\
$m_{Mn}$ & 2.71     &    2.70           &   2.72  &   2.73 \\
$d_{\parallel}^{Mn-Mn}$  & 3.682 & 3.658 & 3.681 & 3.677 \\
$d_{\perp}^{Mn-Mn}$  & 3.675 & 3.677 & 3.641 & 3.666 \\
$d_{\parallel}^{Mn-O}$  &  1.885 & 1.870 & 1.887 &  1.885\\
$d_{\perp}^{Mn-O}$  & 1.877 & 1.878 & 1.857&  1.875 \\
$\angle_{\parallel}$  & 155.23 & 155.99  & 154.77 & 154.51 \\
$\angle_{\perp}$      & 156.49 & 156.36  & 157.39 & 155.67 \\
\hline \hline
&&\textbf{Strained bulk} &&\\
\hline
Magnetic order	& GAF& CAF & AAF & FM  \\
\hline \hline
$\Delta E$  & 0 & 14 & 2 & 23\\
$m_{Mn}$ & 2.787 & 2.810 & 2.82 & 2.881\\
$d_{\parallel}^{Mn-Mn} $ & 3.867 & 3.867&3.867&3.867\\
$d_{\perp}^{Mn-Mn}$  & 3.585 &3.591&3.585&3.593\\
$d_{\parallel}^{Mn-O}$   & 1.973 & 1.973 & 1.975 & 1.975 \\
$d_{\perp}^{Mn-O} $ & 1.849  & 1.856 & 1.850&  1.858 \\
$\angle_{\parallel} $ & 157.01 & 157.13 & 156.60 & 156.40 \\
$\angle_{\perp}     $ & 151.52 & 150.62 & 151.45 & 150.3 \\
\hline \hline				
\end{tabular}
\label{tab-CMO-structural}
\end{table}

In the slab calculations, we focus on the symmetric MnO$_2$ terminated surfaces as it was found that it is more energetically stable than the CaO-terminated one.~\cite{apl_imbrenda} Moreover, in this case the Mn octahedral environment is truncated which could lead to interesting properties. To simulate the thin film grown on SrTiO$_3$, the in-plane lattice parameters are fixed to the one of the fully optimized SrTiO$_3$ within the LDA, (strained-slab). To separate the influence of low dimensionality and the tensile strain we compare the  results obtained for the strained-slab with the ones corresponding to a free standing CaMnO$_3$ thin film, from now on relaxed-slab. In both cases, all the internal coordinates are allowed to relax.

\begin{table*}[ht!]
\caption{Mn-O bond lengths (in \AA) and Mn-O-Mn bond angles (in deg), defined in Fig.~\ref{CMO_surface}, of the relaxed and strain slabs.}
\label{tab-CMO-surface}
\begin{tabular}{@{}|c|cccccccc|}
\hline
Case			& d$_1$&d$_2$ & d$_3$ & d$_4$ & d$_{Mn_1-Mn_3}$ &  d$_{Mn_2-Mn_3}$ & $\angle_{Mn_3-O_7-Mn_2}$ & $\angle_{Mn_1-O_3-Mn_4}$ \\
\hline
relaxed-slab GAF  & 1.87 & 1.92 & 1.82 & 1.86 & 3.60 & 3.76 & 166.0 & 154.0 \\
strained-slab AAF & 2.00 & 2.29 & 1.79 & 1.80 & 3.66 & 4.08 & 170.5 & 151.0\\	
\hline				
\end{tabular}
\end{table*}

\begin{figure*}[ht!]
\includegraphics[width=1.5\columnwidth]{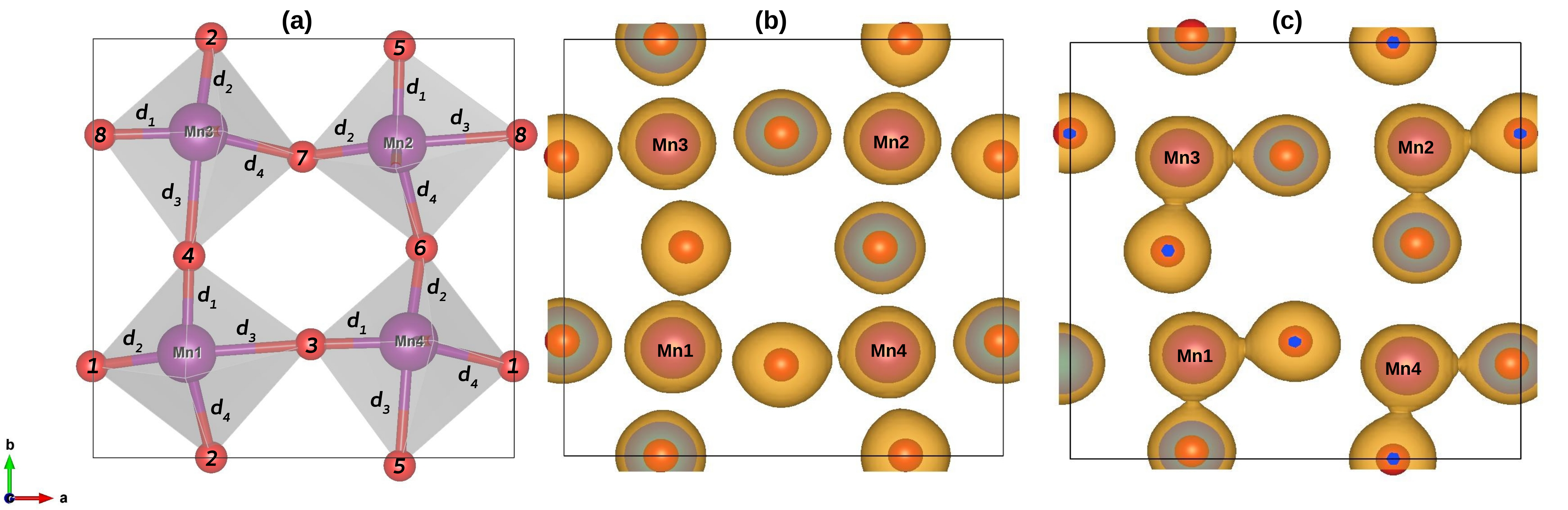}
\caption{(Color online) (a) Top view of the MnO$_2$ surface layer of the strained-slab. The defined distances $d_i$ are shown in Table~\ref{tab-CMO-surface}. (b) and (c): Charge density distribution of
the surface layers of the relaxed-slab and the strained-slab, respectively (top views). The isosurface
value of the charge density was set to 0.135 e/\AA$^3$.}
\label{CMO_surface}
\end{figure*}

As previously mentioned, tensile strain induces changes in the Mn-O bond-lengths and in the  Mn-O-Mn bond angles. In the MnO$_2$ terminated surface the absence of the apical oxygens  introduces structural distortions around the transition metal atom, as clearly seen from Table~\ref{tab-CMO-surface} and Fig.~\ref{CMO_surface}. Due to the octahedral tilting, the presence of this surface breaks the in-plane Mn-Mn bond-lengths equivalency, present in the bulk geometry, into two distinct bond-lengths and the same happens with the in-plane Mn-O-Mn angles. These distortions are present in the relaxed and in the strained slabs but are larger in the strained case, as can also be inferred from Figs.\,\ref{CMO_surface}(b) and (c), where we show the same isovalue of the charge density at the corresponding surfaces of the two slabs.\\
From our total energy calculations we find that the GAF configuration remains being the ground state magnetic configuration for the relaxed-slab but for the strained one, the magnetic ground state corresponds to the AAF case (see Table~\ref{tab-CMO-slab}).

\begin{table}[ht!]
\caption{Energies (in meV/Mn atom.) of the relaxed $Pnma$ and strained thin films for the different magnetic configurations.}
\label{tab-CMO-slab}
\begin{tabular}{@{}ccccc}
Structure			& GAF& CAF & AAF & FM  \\
\hline
relaxed-slab      & \textbf{0}  &   4  &  18  &  36\\
strained-slab     &     4       &  14  &   \textbf{0}  &  16\\	
\hline				
\end{tabular}
\end{table} 

\begin{figure*}[ht!]
\includegraphics[width=.75\columnwidth]{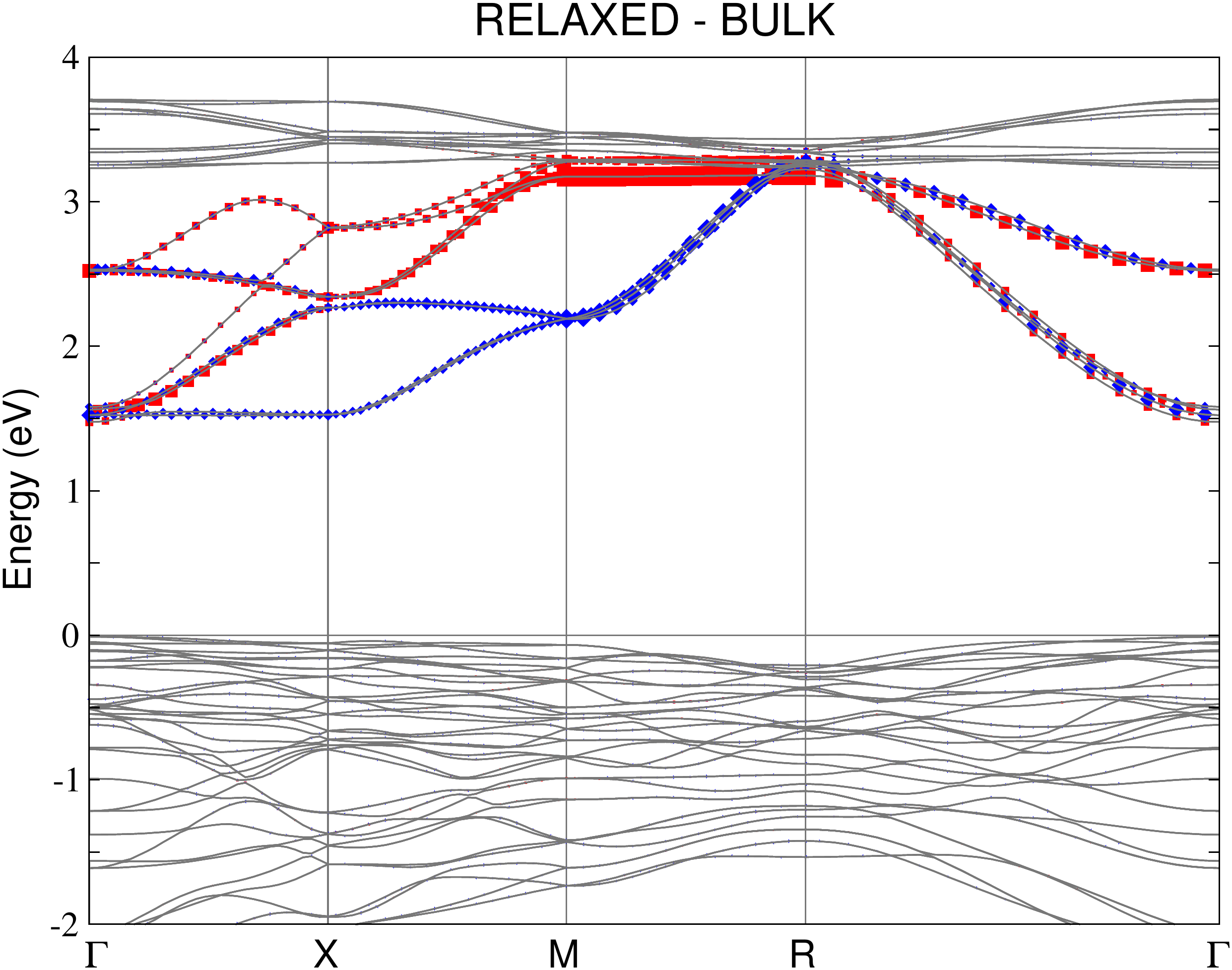}
\includegraphics[width=.75\columnwidth]{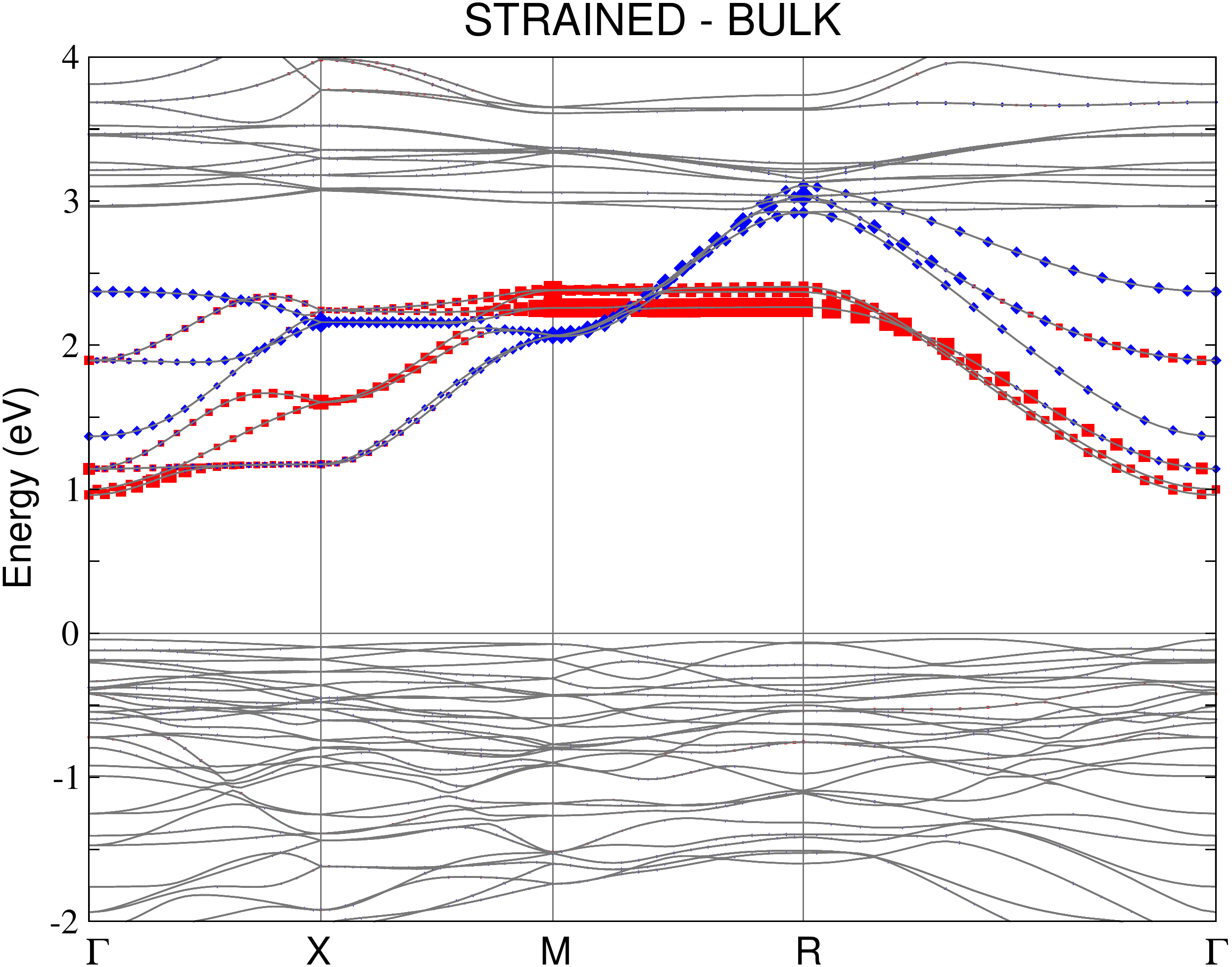}
\includegraphics[width=.75\columnwidth]{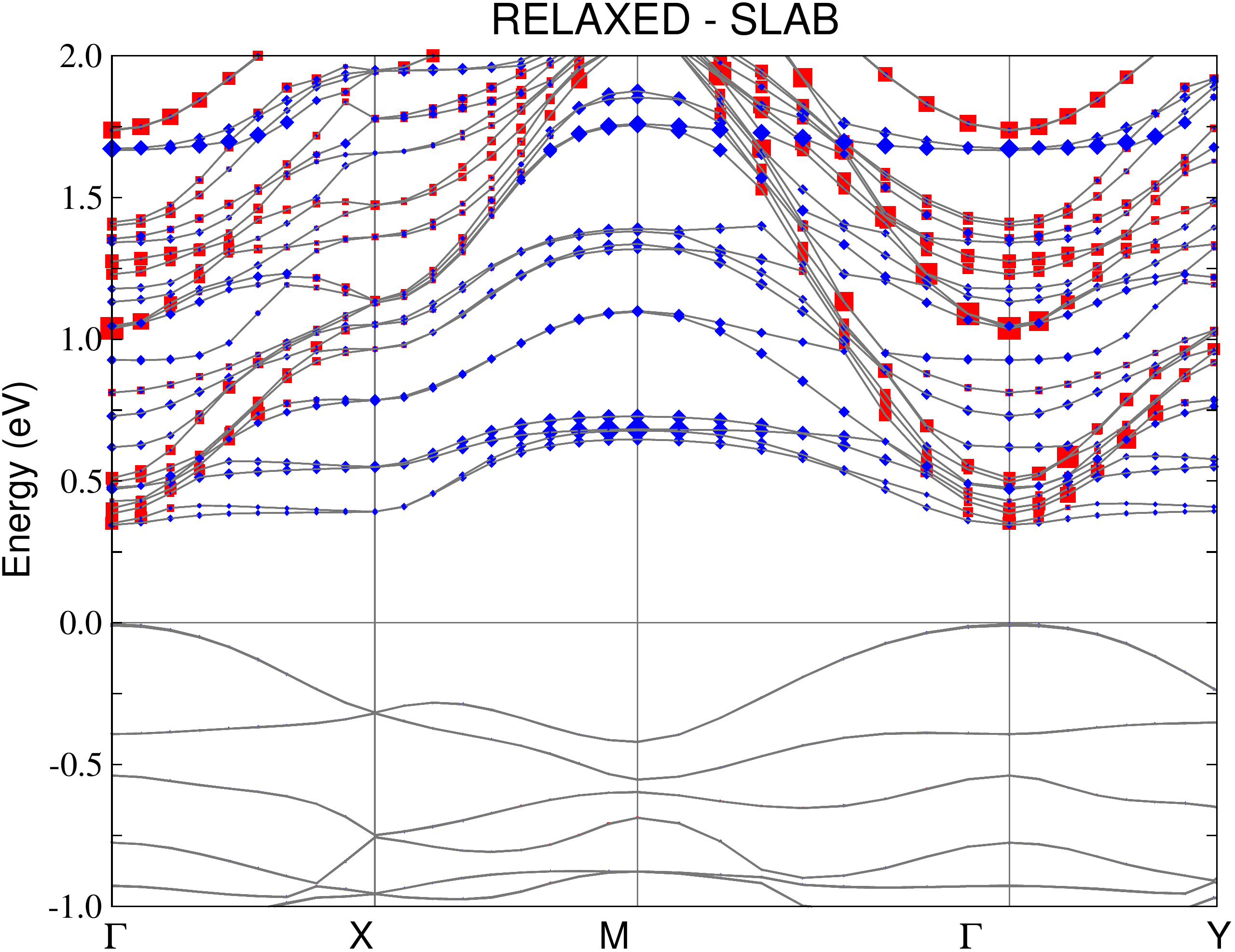}
\includegraphics[width=.75\columnwidth]{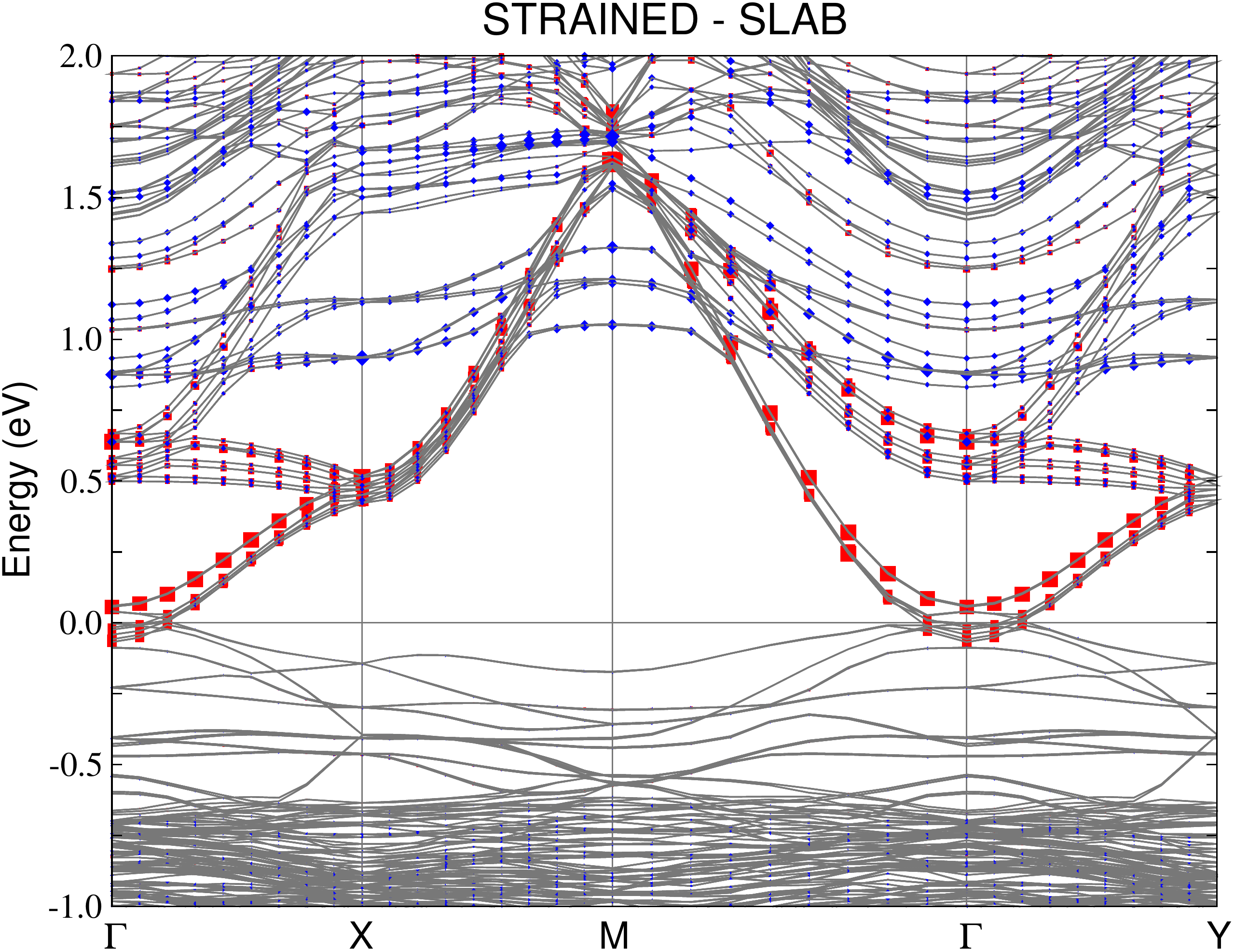}
\caption{(Color online) Electronic bandstructure corresponding to the relaxed and strained bulk CaMnO$_3$ (left and right upper panels, respectively) and the relaxed and strained slabs (left and right bottom panels, respectively). The different $e_g$-orbitals of the Mn atoms are highlighted: red filled squares correspond to $d_{x^2-y^2}$-character and blue filled diamonds to $d_{z^2}$-character. The O-$2p$ orbitals hybridized with the Mn-$3d$ states are localized in the range of energies between -5 eV and the Fermi level. The occupied $t_{2g}$-orbitals lay below the chosen energy range.} 
\label{CMO-slab_bands}
\end{figure*}

\begin{figure}[ht!]
\includegraphics[width=1.\columnwidth]{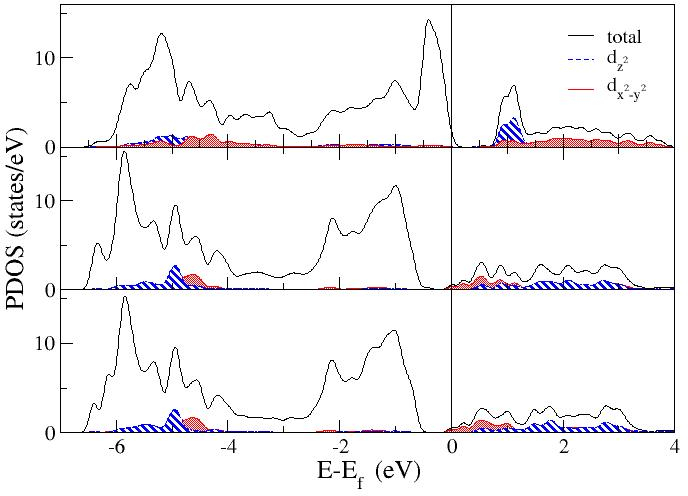}
\caption{(Color online) Spin up partial density of states (PDOS) of the surface (top), third (middle) and central (bottom) MnO$_2$ layers of the strained-slab. Black lines: total DOS of each layer, dashed blue lines: PDOS of each layer projected onto the $d_{x^2-y^2}$-symmetry and red lines: PDOS of each layer projected onto the $d_{z^2}$-symmetry. The occupied $t_{2g}$-orbitals are localized in the range of energies between -6.5 and -3.5 eV.} 
\label{pDOS}
\end{figure}

In Fig.~\ref{CMO-slab_bands} we show the bandstructures of the relaxed and strained bulks of CMO (upper panels) and of the relaxed and strained slabs (lower panels) in their ground state magnetic configurations, namely GAF for the bulk cases and the relaxed-slab and AAF for the strained-slab. The $d_{x^2-y^2}$ and the $d_{z^2}$ orbital characters of the Mn atoms are highlighted (see color-coding in the figure's caption). The valence bandstructure of the relaxed-bulk case is consistent with the experimentally reported one.~\cite{Jung1997} As already mentioned, the Mn $d_{x^2-y^2}$ becomes the lowest empty orbital relative to the $d_{z^2}$ orbital due to the tensile strain (see upper panels of Fig.~\ref{CMO-slab_bands}). On the other hand, the band gap of the relaxed-slab becomes narrower as compared to the one corresponding to the relaxed bulk. As it can be seen from the lower-right panel of the figure, these two effects, namely, tensile strain and low dimensionality, sum up and the Mn $d_{x^2-y^2}$ orbitals become slightly occupied, while the $sp$ orbitals of the surface oxygens, which are totally occupied in the bulk, move up becoming sligthly empty. A deeper analysis of this charge redistribution can be obtained from the partial densities of states (PDOS) depicted in Fig.\,\ref{pDOS}, where we separate the contributions of different MnO$_2$ planes. From the total density of states projected onto each layer, we can see that the oxygens located at the surface are the ones which loose part of their charge, and this charge is redistributed in the Mn atoms located in inner layers, and begin to occupy the $d_{x^2-y^2}$ orbitals. The self-doping electrons of this $e_g$ states,  allow the presence of the FM double exchange and the AAF magnetic structure with more FM pairs becomes more stable. 
This type of coupling gives rise to a net magnetic moment per Mn atom, contrary to what happens in a GAF structure. In the GAF configuration, there is an in-plane compensation of the magnetic moments as the Mn atoms are AFM coupled, regardless the low dimensionality, i.e. the absolute values  of the moments are different from plane to plane, but compensated within each plane. In the AAF configuration, due to  low dimensionality, the absolute values of the magnetic moments differ from plane to plane, but in this case there is not a net compensation within each plane. To give a rough estimation of the net magnetic moment of the ultrathin film, we take into account that an ultrathin film experimentally grown might have an odd or an even number of MnO$_2$ layers, or, zones where there are an odd number of MnO$_2$ layers and others with an even number, as the experimental growth is typically characterized by a certain roughness. The mean value of the magnetic moment taking into account our 9-layers ultrathin film (odd number of MnO$_2$ layers), leads to a net magnetic moment of 0.31$\mu_B$ per Mn atom. If we consider an even number of MnO$_2$ layers, the value of the net magnetic moment is 0.01 $\mu_B$. Therefore, the mean value between the two options (odd or even number of MnO$_2$ layers) yields a Mn uncompensated magnetic moment  within these two values, of around $\mu_B(Mn) \sim 0.16$\,$\mu_B$. This could be the case for an ultrathin film with a  one  MnO$_2$ monolayer roughness. \\
Summarizing, the tensile strain induces a magnetic transition in CaMnO$_3$ ultrathin films. Coupled with this magnetic phase transition we find an electronic transition from an insulator to a metal, as it can be inferred from the right-bottom panel of Fig.~\ref{CMO-slab_bands} and Fig.~\ref{pDOS}. 
As we will show in the next section, this tensile strain is corroborated by means of HRSTEM-HAADF analysis, in which atomic distances are resolved and can be measured (see Fig.~\ref{SEM}).

\section{Experimental C\MakeLowercase{a}M\MakeLowercase{n}O$_3$ ultrathin film grown on S\MakeLowercase{r}T\MakeLowercase{i}O$_3$ }
\label{Exp-details}
The experimental measurements are performed on CMO ultrathin films of 3nm-thickness grown by Pulsed Laser Deposition (PLD) using a Nd:YAG solid state laser onto  (001) SrTiO$_3$ (STO) single-crystalline substrates.  The growth of the films is monitored by in-situ Reflected High Energy Electrons Diffraction (RHEED). During the films growth, the substrates are heated to  700$\degree$C and the oxygen partial background pressure kept at 0.3mbar. After deposition, the samples are cooled to room temperature at an oxygen pressure of 400mbar in order to reduce oxygen vacancies. The crystalline quality,  structure and  stoichiometry of the films are probed by RHEED, high-resolution scanning transmission electron microscopy (HRSTEM) coupled  with high angular dark field detector (HAADF) and Energy Dispersive X-Ray Analysis (EDX). The HRSTEM-HAADF measurements are performed in a FEI Titan G2 at 300keV probe corrected.  The roughness of the films is examined by Atomic Force Microscopy (AFM). In addition, the magnetic properties of the films are studied using a SQUID magnetometer (Quantum Design MPMS) performing magnetization loops at low temperature in two geometries, i.e. magnetic field applied in-plane (IP) and out-of-plane (OOP) of the samples.\\

\begin{figure}[ht!]
\includegraphics[width=0.7\columnwidth]{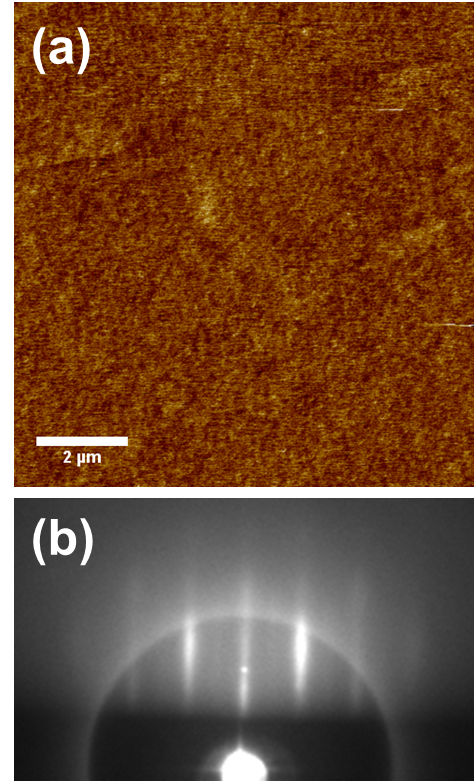}
\caption{(a) AFM image and (b) RHEED pattern of a 3nm-thick CMO thin film .}
\label{afm}
\end{figure}

\begin{figure}[ht!]
\includegraphics[width=1\columnwidth]{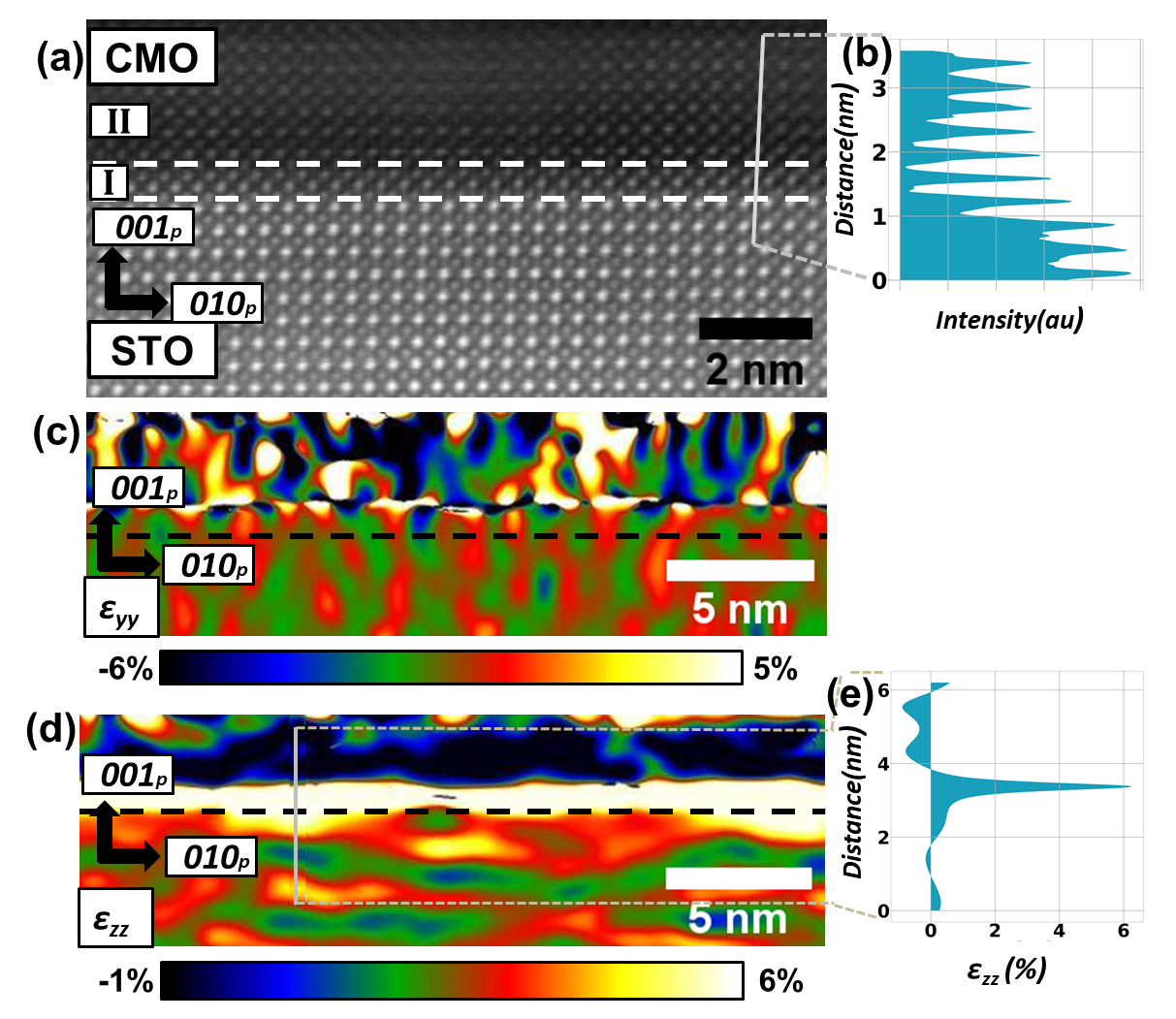}
\caption{(color online) (a) Cross sectional HRSTEM-HAADF image of the interface CMO/STO along [100]$_p$ zone axis, the white dotted lines indicate the two regions of the CMO thin film. (b) Intensity profile along [001]$_p$ direction to measure the interplanar distances. GPA strain analysis maps of the strain components (c) $\epsilon_{yy}$ parallel to the interface, and taking the pole diffraction [010]$_p$, and (d) $\epsilon_{zz}$ perpendicular to the interface and taking the [001]$_p$ pole diffraction, respectively. The black dotted line indicates the interface between the substrate and the CMO thin film.(e) $\epsilon_{zz}$ profile along the [001]$_p$ characterized by a sharp expansion at the interface}
\label{SEM}
\end{figure}

The RHEED patterns of the films are characteristic of flat surfaces, in agreement with results obtained by AFM  which show a RMS roughness lower than 0.5nm (Fig \ref{afm}).
The stacking of CMO onto the STO as a continuation of  a pseudocubic cell is analysed in HRSTEM-HAADF images, where the parameter $a_p$ is $\frac{a}{\sqrt{2}}$ and c,  the out of plane parameter is similar to $a_p$ (see Fig.\ref{SEM} (a)). Two different CMO regions, CMO I and CMO II, are clearly distinguished from the images,   characterized by a sharp discontinuity of their lattice parameters. The lattice parameters are calculated from the HRSTEM-HAADF images using the Digital Micrograph DM3 software averaging the atomic distance for different columns and rows.
~\cite{gpa_paper4} Closer to the STO, CMO I is 1nm thick and is almost cubic with an average lattice parameter  of $a_p$ = (3.93 $\pm$ 0.02)\AA. A progressive relaxation of the CMO lattice is observed in the CMO II region (see Fig.~\ref{SEM}a.) which is tensile strained in the plane of the films. 
The interplanar distances between atom columns are measured by Digital Micrograph from the intensities of the STEM-HAADF detector (Fig.~\ref{SEM}b). Along this last zone, the in plane (IP) lattice parameter, $a_p$, relaxes from (3.88 $\pm$ 0.02)\AA~ to (3.77 $\pm$ 0.02)\AA~ while the out of plane (OOP) parameter, \textit{c}, varies from (3.78 $\pm$ 0.02)\AA~ to (3.68 $\pm$ 0.02)\AA. The c/a ratio increases smoothly from 0.974 to 0.976. It is worth to notice that the crystalline structure of the CMO films is notably distorted with respect to the bulk one, with an expanded unit cell  and critically defined by the substrate induced biaxial strains and distortions.\\
Complementary studies are performed to analyze the strain across the samples using the Geometric Phase Analysis (GPA) software for Digital Micrograph on the STEM-HAADF images.~\cite{gpa_paper,gpa_paper2,gpa_paper3} The results obtained for the IP strain (See Fig.\ref{SEM} (c)) and the OOP strain (See Fig.\ref{SEM} (d)) clearly show the two regions observed in the STEM-HAADF images. In the case of the IP strain the CMO film and the substrate present similar colors near the interface, which indicates similar strain at both sides of the interface. Above this region, the CMO film presents a more complex strain field, with bright colors which indicate regions with positive strain and dark colors which indicate regions with negative strain. This type of strain field is common in columnar or mosaic-like growth. The OOP strain field shows a bright region near the interface which indicates a positive strain and a sharp transition to a relaxed structure. This behaviour can be observed from the $\epsilon_{zz}$ profile along the [001]$_p$ direction in which a positive strain of 6\% is observed just above the interface (See Fig.\ref{SEM} (e)). The stoichiometry of the samples presents an anomalous behavior and has a direct correlation with the results obtained for the structure of the samples. Near the interface, we observe a major concentration of Mn ions and a deficiency of Ca ions, specially at the first three atomic layers, in agreement with the strained region observed in the previous analysis. This excess of Mn ions might indicate the presence of Mn oxides at the interface. After this region, we observe the expected stoichiometry for CMO. 

\begin{figure}[ht!]
\includegraphics[width=0.9\columnwidth]{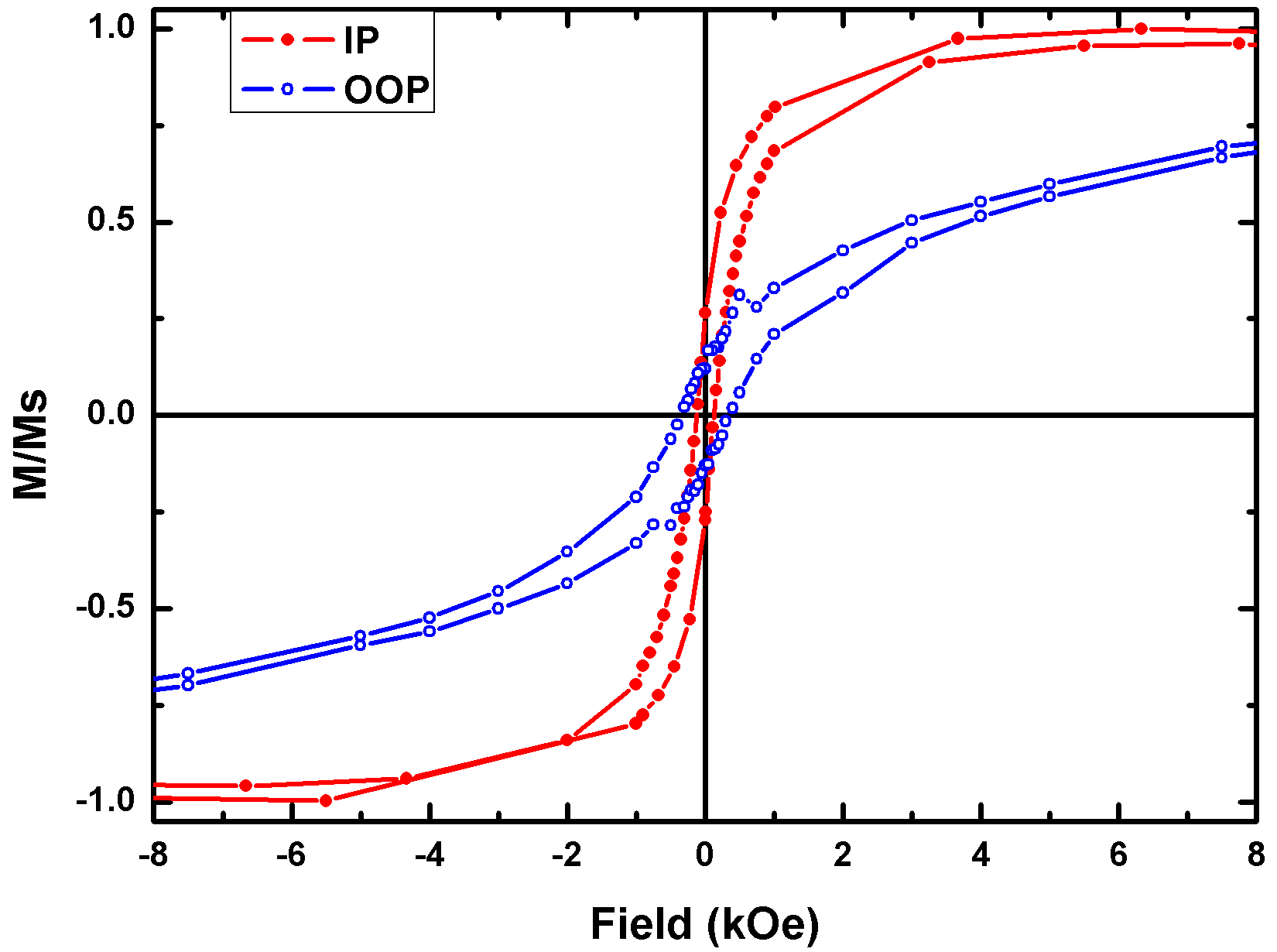}
\caption{(\textcolor{red}{$\CIRCLE$})In-plane and (\textcolor{blue}{\textbf{$\Circle$}}) out-of-plane magnetization loop for the CMO thin film at 10K.}
\label{mag}
\end{figure}

Magnetization loops are obtained in the OOP and the IP geometries of the applied magnetic field at 10K. The IP loops present hysteresis with a  saturation magnetization (M$_s$) of (120$\pm$25)emu/cm$^3$, which gives an average magnetic moment per Mn atom of $\mu_{Mn}$ = $(0.67\pm 0.14)\mu_B$, and a remanence magnetization, (M$_r$), of about 20\% of the M$_s$ (Fig. \ref{mag}). The IP magnetization saturates above 4kOe. A comparison between the OOP and IP loops indicates clearly that CMO has an easy-plane anisotropy.  

As already mentioned, in bulk, CMO presents weak ferromagnetism, characterized by a small M$_s$ of 0.04  $\mu_B$/Mn at 10K and a GAF configuration.\cite{Zeng1999,pr_macchesney,fujishiro_cmo} The high M$_s$ observed in our samples differs from the bulk value, showing more resemblance to a ferromagnetic coupling than to a weak ferromagnetic coupling, but the small M$_r$ cannot be explained with a pure ferromagnetic coupling. In other words, the high M$_s$ value cannot be explained just by a weak ferromagnetic coupling. The origin of these results is not completely clear and can have different possible explanations. The STEM-HAADF images and the GPA analysis show that the CMO thin film is heavily influenced by the substrate, changing the crystal and the electronic structure and also the stoichiometry at the interface. 

\section{Discussions and Conclusions}
In this work we analyze the effect of high tensile strain and low dimensionality on the magnetic and electronic properties of CaMnO$_3$ ultrathin films epitaxially grown on SrTiO$_3$ substrates, both, computationally and experimentally. 
From he computational simulations,  we find that the combination of both effects yields a change in the magnetic order, namely, the GAF structure of bulk CMO changes to an AAF configuration. This should lead to the appearance of a ferromagnetic signal, which would not be present if the magnetic structure were GAF. This signal should come from non-compensated ferromagnetic MnO$_2$ superficial layers.
Remarkably, the FM component of the magnetic structure is stabilized, not as a consequence of a direct electron doping, as happens when substituting the divalent Ca by trivalent elements or in the presence of O vacancies, as reported in the past~\cite{La-doped1,La-doped2,Molinari2014}, but rather because of a self-doping charge redistribution due to, both the strain and the absence of apical oxygens.  On the other hand, we also find that the mentioned  magnetic phase transition is coupled to an electronic transition from an insulator to a metal. 

When the ultrathin CaMnO$_3$ films are experimentally grown on an SrTiO$_3$ substrate, the appearance of a ferromagnetic hysteresis loop is, indeed, observed, together with an easy plane due to shape anisotropy. The experimentally obtained net magnetic moment per Mn atom $(0.67\pm 0.14)\mu_B$ is bigger than the theoretically estimated value, indicating that other ingredients are present in the grown samples that contribute to the net magnetic moment. The Mn excess at the interface, observed in the HRSTEM-HAADF analysis, might be behind the increased  ferromagnetic signal. Another minor contribution  could be due to the rearrangement of the oxygen atoms at the interface, giving rise to a distortion of the interfacial oxygens' octahedra,  yielding an enhanced Dzyaloshinskii-Moriya interaction and, thus, a larger weak ferromagnetic component.~\cite{Spaldin2012,Graf2018}  The  oxygen octahedral distortion (OOR)  is a consequence not only of the strain imposed by the substrate, but also of the discontinuity present at the interface which changes from  $a^0a^0a^0$ (in Glazer's notation) in STO to $a^-a^-c^+$ in CMO, similar to what was found in other heterointerfaces.\,\cite{NatMat2016,Triscone2019,tensiones1,tensiones2,tensiones3} Nonetheless, these distortions might be important only in the first two layers close to the interface, as it was reported in Ref.\,\onlinecite{2014-mismatch}. Therefore,
we consider the effect of the symmetry mismatch as being less relevant with respect to the strain, which extends all over the ultrathin thickness.\,\footnote{To give a first glance in the effect of the symmetry mismatch introduced by the presence of the substrate, we perform further calculations in the CaMnO$_3$/SrTiO$_3$ slab, where we consider nine layers of CMO grown over two layers of STO. We constrain the first layer of STO to the LDA optimized corresponding one, and allow the other STO layer as well as all the CMO layers to relax their internal coordinates. We find that the AAF magnetic configuration is still the most stable one, and that the magnetic moments we are reporting are still consistent with this calculation.} Besides, in our experimental
counterpart, no interdiffusion is observed.

It is worth mentioning that, to the best of our knowledge, it is the first time that magnetic properties of ultrathin CaMnO$_3$ films under high tensile strain are investigated both experimentally and theoretically. Even though our computational and experimental findings are not totally coincident, interesting properties emerge from the combination of low dimensionality and strain effects. In view of the potential technological applications of this compound in antiferromagnetic nano-spintronics, a deeper understanding is a must. From the computational side this understanding could be achieved by introducing, in the future, defects such oxygen vacancies, the formation of a Mn-oxide buffer layer or the explicit presence of the SrTiO$_3$ substrate. From the experimental side, the improvement in the control of interfaces and the absence of oxygen vacancies should lead to a better comprehension of the different physical properties triggered by strain and surface effects.

\section*{Acknowledgments}
This work was partially supported by PICT-2016-0867 of the ANPCyT, Argentina, and by H2020-MSCA-RISE-2016 SPICOLOST Project Nº 734187.  Authors acknowledge the Laboratory of Advanced Microscopy at Instittute of Nanoscience of Arag\'on for offering access to their instruments .

\bibliography{biblio}
\end{document}